\documentclass[12pt]{article}
\usepackage[cp1251]{inputenc}
\usepackage[english]{babel}
\usepackage{amssymb,amsmath}
\usepackage{graphicx}

\title{ On the nucleons motion in the nucleus as
being due to realistic inter-nucleonic forces}

\author{Yu.P.Lyakhno}

\begin {document}

\maketitle

\begin{center} {\it National Science Center "Kharkov Institute of
Physics and Technology" \\ 61108, Kharkiv, Ukraine}
\end{center}

\begin{abstract}

This paper deals with the possible motion of nucleons in the
nucleus, which is due to realistic inter-nucleonic forces. This
approach provides new or more substantiated conclusions about the
nuclear structure than those based on the effective interaction of
nucleons, while the shell model of the nucleus may lead to
questionable conclusions regarding the nuclear structure and nuclear
reaction mechanisms.

\vskip20pt


\end{abstract}

\section{Introduction}
High-accuracy data on two-nucleon forces acting in the nucleus have
been obtained in recent years. These data enable one to gain more
reliable information on the nuclear structure and the nuclear
reaction mechanisms than the information derived on the basis of
effective nucleon-nucleus interaction. For illustration, figure 1
shows the CD-Bonn data for the realistic {\it NN} potential
\cite{1,2}. They were derived from the phase analysis of the
experimental data on elastic ({\it p,p}) and ({\it n,p}) scattering
in the energy range up to 350 MeV for the total momentum $J\le$4 of
the {\it NN} system (notation: $^{2S+1}L_J$, {\it S} -spin, {\it L}
-orbital momentum of two-nucleonic system). The positive phase value
corresponds to inter-nucleonic attraction, while the negative phase
value corresponds to nucleon-nucleon repulsion. As the nucleon
scattering energy E$_N$ tends to zero, the $^3S_1$ phase goes to
$180^0$, and this corresponds to the bound state of the {\it np}
system, i.e., the deuteron. It is necessary to point out that
nucleon modification is possible in the nucleus \cite{3}. However,
if this nucleon modification in the nucleus does exist, then it will
also occur at two-nucleon scattering, and consequently, will be
taken into account phenomenologically, too.

A distinctive feature of the inter-nucleonic forces is that they
depend not only on the inter-nucleon distance {\it r}, but also on
the quantum configuration of the nucleon system, which is determined
by the orbital momentum {\it L}, spin {\it S} and isospin {\it T} of
the system. The dependence of the {\it NN} potential on the quantum
configuration of the two-nucleon system may be more considerable
than the dependence on the distance {\it r}. Besides, 3{\it N}
forces are also at work in the nucleus \cite{4}, \cite{5}.
Therefore, the nuclear wave function, which is dependent only on the
distance {\it r}, may be inconsistent with reality.
\begin{figure}[h]
\noindent\centering{
\includegraphics[width=90mm]{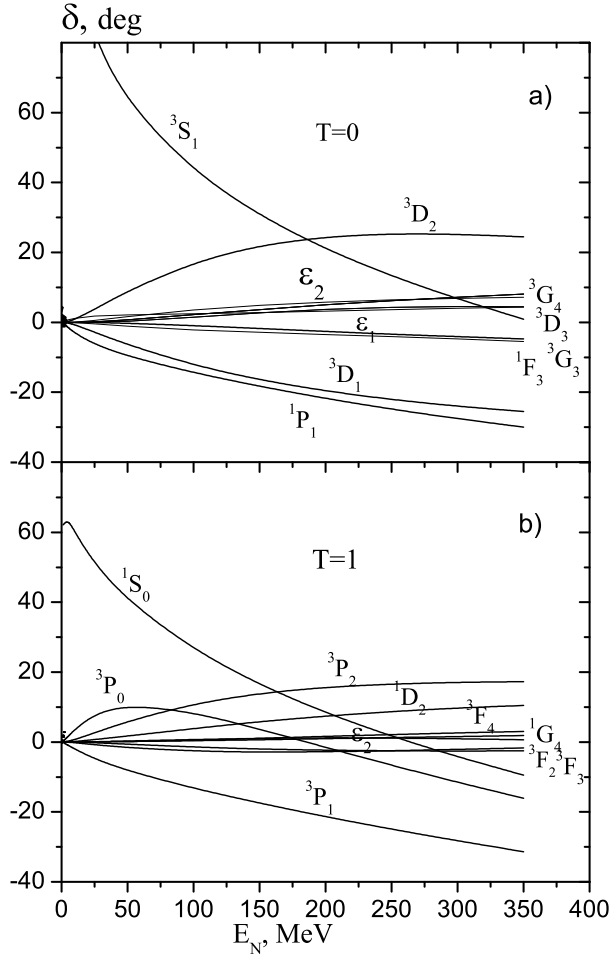}
} \caption{Phases $\delta$ and mixing parameters $\epsilon$ of {\it
NN} scattering: a) ({\it n,p}) scattering with isospin  {\it T}=0
and b) ({\it n,p}), ({\it p,p}) and ({\it n,n}) scattering with {\it
T}=1 in the CD-Bonn nucleon-nucleon potential. Insignificant
differences between phases in the {\it T}=1 states are not shown in
the figure.}
\end{figure}

\section{ Nucleon pairing and clustering of nucleus }

Using the {\it NN} potentials from refs. \cite{1} and \cite{6}, and
also the 3{\it N} forces UrbanaIX \cite{4} and Tucson-Melbourne
\cite{5}), the authors of ref. \cite{7} have calculated the binding
energies, the probability of states with nonzero orbital momenta of
nucleons, and the nucleon momentum distribution in the $^4$He
nucleus. The calculations of the nucleon momentum distribution and
other mentioned parameters are in good agreement with the available
experimental data. This distribution is shown in Fig.2 as a function
of relative energy E$_N$ of two nucleons.
\begin{figure}[h]
\noindent\centering{
\includegraphics[width=130mm]{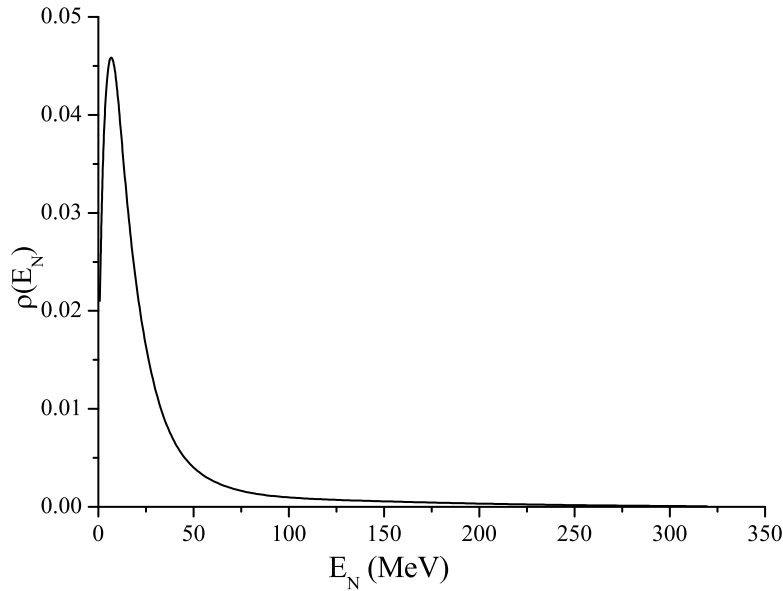}
} \caption{Nucleon energy distribution in the $^4$He nucleus. The
functions are normalized to $\int f(E_N)dE_N$=1.}
\end{figure}
The comparison between figures 1 and 2 suggests the conclusion that
for the most part of time the nucleons stay in the states with zero
orbital momenta. The neutron-proton pairing in the {\it T}=1 state
and also neutron-neutron, proton-proton pairings, take place in the
$^1S_0$ state, i.e., in the state with antiparallel spins, while the
pairing of  neutron with proton at {\it T}=0 occurs in the $^3S_1$
state. The pairing takes place in a wide range of relative energies
of nucleons, i.e., is dynamic. The paired nucleons are bosons, and
therefore, the Pauli principle doesn't forbid the basic part of the
nucleons of the nucleus to be in the paired states.

The attraction between nucleons having zero orbital momenta can lead
to nuclear clusterization. The 3{\it N} forces also have an effect
of attraction. This makes an additional contribution to
clusterization of the nucleus. The formation of several similar
three-nucleon clusters can be suppressed in accordance with the
Pauli principle. However, after attraction of the fourth nucleon,
the three-nucleon cluster would not impede a further clusterization
of the nucleus. For example, the paired neutron from the
neutron-proton pair in  the $^3S_1$, {\it T}=0 state  takes on  the
other neutron in $^1S_0$,{\it T}=1 state, while the proton of the
mentioned pair joins with the other proton in the same state
(fig.3a); that leads to cluster formation in the $^1S_0$ state.
Similarly, a cluster can appear due to $^1S_0$,{\it T}=1 couplings
(fig.3b). Also, the clusters can form from four neutrons or four
protons by means of $^1S_0$,{\it T}=1 couplings (fig.3c). The
possibility of formation of more complex nucleonic clusters is not
excluded, in particular, cluster from 8 neutrons. It is possible,
that this is connected to the fact that nuclei $^{40}$Ca and
$^{48}$Ca are magic.

\begin{figure}[h]
\noindent\centering{
\includegraphics[width=100mm]{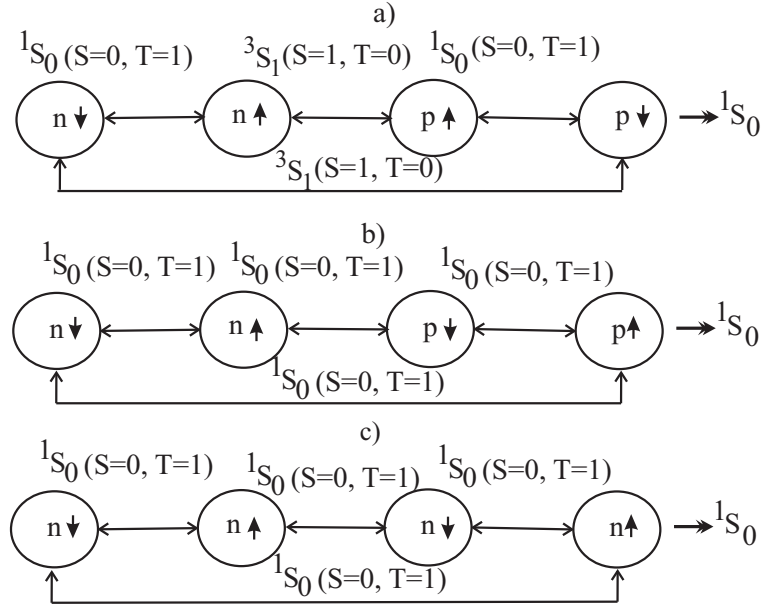}
} \caption{Schemes  of $^1S_0$ $\alpha$-cluster formation.}
\end{figure}

Pairing of nucleons and clusterization of nuclei are confirmed by
numerous experimental data. Thus,  the nucleus represents
essentially the boson system, and this makes  the  application  of
the shell model to the nucleus unreasonable.

\section{ Non-central strong interaction of nucleons }

Non-centrality of strong interaction leads to the fact that in the
process of intranuclear motion the nucleon spin-flip may take place.
For example, with spin-flip of one nucleon in the $^4$He  nucleus,
the nuclear spin will take on the {\it S}=1 value. At that,
according to the laws of conservation of total momentum and parity
of the nucleus $^4$He, we have $J^{\pi}=0^+$, the total orbital
momentum of nucleons would have to change and take on the {\it L}=1
value. That leads to the emergence of states with non-zero orbital
momenta of nucleons in the lightest nuclei. The laws of conservation
of total momentum, parity, and also, the Pauli principle, permit
only two values of orbital momentum of nucleons in the deuteron,
viz., {\it L}=0 and {\it L}=2.  In the rest of the nuclei with
A$\ge$3 there are infinitely many states with non-zero orbital
momenta \cite{7}. However, with increase in the orbital momentum of
the nucleon the probability of the mentioned state decreases (see
fig.1). Therefore, later on, the nucleon will return to its original
state, provided that it remained unoccupied by another nucleon.
Similarly, spin-flips of two nucleons may occur, and the $^4$He
nucleus will appear in the {\it S}=2 and {\it L}=2 state. The
probabilities of states with non-zero orbital momentum of the
nucleons of the lightest nuclei have been calculated in \cite{8}.

When constructing the effective interaction of nucleons, the curves
in fig.1 are averaged, including the 3{\it N} forces, which can also
lead to the nucleon spin-flip. As a result, the nuclear shell model
predicts the total momenta {\it J} and the spins  {\it S} of all
even-even nuclei to be zero. The experimental data obtained from the
studies of the $^4$He($\gamma,p)^3$H and $^4$He($\gamma,n)^3$He
reactions with emission of one nucleon, show that in these reactions
the multipole transitions with the spin {\it S}=1 of the final-state
of the particle system take place. Their contribution is about
$\sim$10$^{-2}$ of the total reaction cross-section. The
experimental information about these {\it S}=1 transitions  can be
obtained, in particular, in the collinear geometry, in which the
contribution of dominant transitions with the spin {\it S}=0 is
absent. In theoretical \cite{9} and experimental \cite{10} works,
the occurrence of {\it S}=1 transitions was attributed to the fact
that the electromagnetic interaction caused a spin-flip of the
hadronic particle system, and that spin-flip was due to the
contribution of mesonic exchange currents (MEC). It should be noted
that the MEC contribution depends on the photon energy \cite{11}.

The calculation \cite{7} based on realistic inter-nucleonic forces
has shown that the ground state of the $^4$He nucleus can be in the
states with non-zero orbital momenta of nucleons, and the spin of
the $^4$He nucleus can take the values  {\it S} = 0, 1 and 2.
Consequently, the transitions with spin {\it S}=1 of the final-state
of the particle system can originate from the initial state of the
$^4$He nucleus with spin {\it S}=1 without spin-flip of the nucleon
in the process of reaction. In this case the ratio of the total
cross-section of {\it S}=1 transitions to the total cross-section of
the reaction can be independent of the photon energy \cite{12}.

The analysis of the experimental data (fig.4) has suggested the
conclusion that, within the statistical error, the ratio of the
reaction cross-section in the collinear geometry to the
cross-section of the electrical dipole transition with {\it S}=0 at
the angle of nucleon emission $\theta_N$ = 90$^0$ ($\nu_p$ and
$\nu_n)$ in the photon energy range 22$\le$E$_{\gamma}$$\le$100 MeV
does not depend on the photon energy (despite the fact that in this
photon energy range, the total cross-section of the reaction
$\sim$15 times). The average $\nu_p$ and $\nu_n$ values in the
mentioned photon energy range were calculated to be $\nu_p$ = 0.01
$\pm$ 0.002 and $\nu_n$ = 0.015 $\pm$ 0.003, respectively. The
calculations took into account the errors in the measurement of the
polar angle of nucleon emission in the mentioned reactions
\cite{12}. The available experimental data are in agreement with the
theoretical calculation \cite{7}, and also, with the assumption that
the {\it S}=1 transitions can originate from {\it P}-states of the
$^4$He nucleus.

Thus, the conclusion about spin flipping during the reaction, made
on the basis of the nuclear shell model, raises doubts. The
conclusions about the nucleon knocking-out from {\it s} - and {\it
p} -shells of the nucleus may also be open to question.

\begin{figure}[h]
\noindent\centering{
\includegraphics[width=100mm]{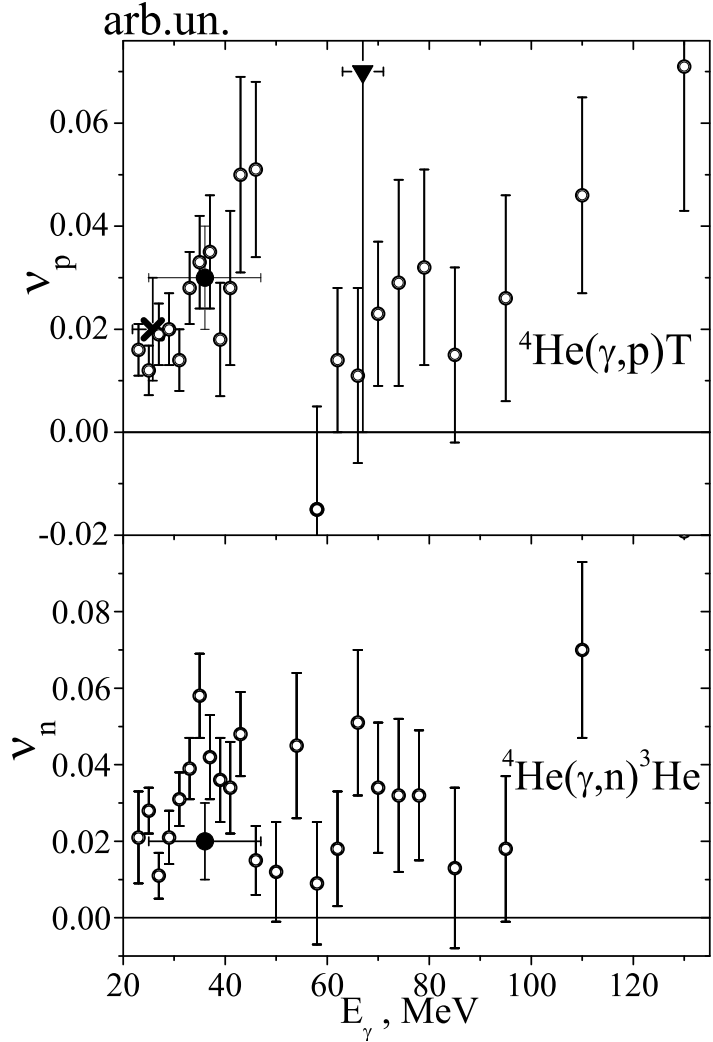}
} \caption{Ratio of the reaction cross-section in the collinear
geometry to the cross-section of the electrical dipole transition
with {\it S} = 0 at the nucleon emission angle $\theta_N$ = 90$^0$.
Closed points: Balestra et  al. \cite{13}; triangle: Jones et al.
\cite{14}; open points: Arkatov et al. \cite{15}; cross: Shima et
al. \cite{16}.}
\end{figure}
The presence of states with non-zero orbital momenta of nucleons in
the lightest nuclei is due to the tensor part of the {\it NN}
potential, and also, 3{\it N} forces. Consequently, similar effects
must be observed unexceptionally in all the nuclei, including their
excited states, too. It can be assumed that with an increase in the
number of nucleons A in the nucleus, the number of nucleons with
flipped spins also increases. This increase for the nucleus with the
number of nucleons A relative to the nucleus with the number of
nucleons A-1 can be estimated from the contribution of {\it D}-wave
to the deuteron wave function, i.e., $\sim$ 5\%. In the general
case, the spin of the nucleus with the number of nucleons A can take
on integer values in the interval 0$\le$ S$\le$ A/2 provided that A
is even, or half-integer values in the interval 1/2$\le$ S$\le$ A/2
provided that A is odd. The total orbital momentum of the nucleons
{\it L} must take the values in accordance with the laws of
conservation of the total momentum and parity of the ground state of
the nucleus or its excitation level.

It can be supposed that in the process of pairing the odd neutron or
the odd proton in odd-odd medium and heavy nuclei generally appears
to be in the states with non-zero orbital momentum. Therefore, in
these nuclei the odd proton and the odd neutron cannot be paired in
the $^3S_1$ state. Perhaps for this reason, only very light odd-odd
nuclei are stable.

\section{ Spin-orbit interaction of nucleons in the nucleus}

The spin-orbit interaction of nucleons leads to an additional
contribution to the potential energy of the nucleus. Within the
framework of the nuclear shell model this energy can be calculated
by the expression:

\begin{equation}
\label {eq13}
 U_{SO}=-b\bigg(\frac{\hbar}{Mc}\bigg)^2\sum_{i=1}^A\frac{1}{r_i}\frac{\partial V_i}{\partial r_i}
 (\vec l_i \cdot\vec s_i)\,,
\end{equation}
where {\it V} is the spherically symmetrical potential, {\it l} is
the orbital momentum,  {\it s} is the nucleon spin. However, to
bring into agreement with the experiment, expression (1) should be
multiplied by the spin-orbit interaction constant of nucleons {\it
b}. For medium and heavy nuclei the constant comes up to $b\sim$10,
and this value increases with increasing A. The origin of the
constant may be attributed to the fact that in the medium and heavy
nuclei the significant number of nucleons is in the spin-flip
states. For example, let us assume that in the $^{208}$Pb nucleus
ten nucleons are in the spin-flip states, i.e., the nucleus spin is
{\it S} = 10. Then the total orbital momentum of the nucleons must
be {\it L} = 10. This can give rise to a substantially higher
contribution of the spin-orbit nucleon interaction than that
predicted by the nuclear shell model. This can be a part of the
reason for the origin of the constant {\it b} of the spin-orbit
nucleon interaction.

\section{ Conclusions}

Two competing processes are at work in the nucleus. On the one hand,
the realistic  inter-nucleonic forces result in nucleon pairing and
nucleus clustering. On the other hand, the non-centrality of the
strong nucleon interaction and the 3{\it N} forces cause spin-flips
of the nucleons, and consequently, the decay of the formed nucleon
pairs and their clusters. The nucleus spin {\it S} and the total
orbital momentum of the nucleons {\it L} are random variables, with
the distribution dependent on the specific nucleus.

Despite a considerable MEC contribution to the total cross section
of the reaction, the contribution of the spin-flip of the hadronic
particle system as a result of electromagnetic interaction may be
suppressed.

The nuclear shell model may lead to doubtful conclusions about the
nuclear structure and mechanisms of nuclear reactions.

The author gratefully acknowledges a fruitful discussion with Dr's
A.F.Khodyachikh and E.A.Skakun.

\end{document}